\documentclass[conference]{IEEEtran}
\IEEEoverridecommandlockouts
\usepackage{cite}
\usepackage{amsmath,amssymb,amsfonts}
\usepackage{algorithmic}
\usepackage{graphicx}
\usepackage{textcomp}
\usepackage{xcolor}
\usepackage{subcaption}
\def\BibTeX{{\rm B\kern-.05em{\sc i\kern-.025em b}\kern-.08em
    T\kern-.1667em\lower.7ex\hbox{E}\kern-.125emX}}

\newcommand{\norm}[1]{\left\Vert#1\right\Vert}

\newcommand{\Real}{\mathbb R}

\begin{document}

\title{BP-DIP: A Backprojection based Deep Image Prior
\thanks{This work is funded by ERC-SPADE and Magnet MDM.}
}

\author{\IEEEauthorblockN{Jenny Zukerman}
\IEEEauthorblockA{\textit{Department of Biomedical Engineering} \\
\textit{Tel Aviv University}\\
Tel Aviv, Israel\\
jennyz@mail.tau.ac.il}
\and
\IEEEauthorblockN{Tom Tirer}
\IEEEauthorblockA{\textit{School of Electrical Engineering} \\
\textit{Tel Aviv University}\\
Tel Aviv, Israel\\
tomtirer@mail.tau.ac.il}
\and
\IEEEauthorblockN{Raja Giryes}
\IEEEauthorblockA{\textit{School of Electrical Engineering} \\
\textit{Tel Aviv University}\\
Tel Aviv, Israel\\
raja@tauex.tau.ac.il}
}

\maketitle

\begin{abstract}
Deep neural networks are a very powerful tool for many computer vision tasks, including image restoration, exhibiting state-of-the-art results. However, the performance of deep learning methods tends to drop once the observation model used in training mismatches the one in test time. In addition, most deep learning methods require vast amounts of training data, which are not accessible in many applications. 
To mitigate these disadvantages, we propose to combine two image restoration approaches: (i) Deep Image Prior (DIP), which trains a convolutional neural network (CNN) from scratch in test time using the given degraded image. It does not require any training data and builds on the implicit prior imposed by the CNN architecture; and (ii) a backprojection (BP) fidelity term, which is an alternative to the standard least squares loss that is usually used in previous DIP works. We demonstrate the performance of the proposed method, termed BP-DIP, on the deblurring task and show its advantages over the plain DIP, with both higher PSNR values and better inference run-time.
\end{abstract}


\section{Introduction}

Image restoration refers to the recovery of an original unknown image from its degraded version, which suffers from defects, such as blur, noise and low resolution. In a linear image restoration problem, the goal is to recover the original image $x^* \in \Real^{n}$ from the degraded measurements 
\begin{equation}
\label{eq:inverse_prob}
    y = Ax^* + e,
\end{equation}
where $e \in \Real^{m}$ is an additive noise and $A \in \Real^{m \times n}$ is a degradation operator. For example, in image deblurring $m=n$ and $A$ is a square ill-conditioned matrix which represents a blur operator that filters the image by a blur kernel. %, e.g. a gaussian or a uniform filter.

Image restoration tasks often involve minimization of a cost function, composed of a fidelity term and a prior term
\begin{equation}
\label{eq:cost_function}
    \min_{x} \ell(x,y) + \beta s(x),
\end{equation}
where $\ell$ is the fidelity term, $s$ is the prior term and $\beta$ is a positive parameter that controls the level of regularization. The fidelity term forces the output image to comply with the observation model, while the prior poses an underlying assumption about the latent image, e.g. that natural images are sharp and free of noise and holes.

Since inverse problems represented by (\ref{eq:inverse_prob}) are usually ill-posed, a vast amount of research has been focused on the prior term $s(x)$. Various natural image priors have been researched. Some of them can be described by explicit and interpretable  functions, e.g. total-variation (TV) \cite{Rudin1992NonlinearTV}, and others, such as BM3D \cite{Dabov2007ImageDB} and (pre-trained) deep generative models \cite{Bora2017CompressedSU}, are more implicit (i.e. cannot be associated with explicit prior functions). %An additional prior is the non-local similarity %(NLS) \cite{Buades2005ANA}, \cite{Glasner2009SuperresolutionFA}.
Many image priors (of both kinds) also exploit the non-local similarity of natural images \cite{Buades2005ANA, Glasner2009SuperresolutionFA}.

The fidelity term, however, has been less researched and is often chosen as the typical least squares (LS) objective. Recently, the authors of \cite{Tirer2019BackProjectionBF} have presented the backprojection (BP) fidelity term, which shows an advantage over the standard LS term on different image restoration tasks, such as deblurring and super-resolution, using priors such as TV, BM3D, and convolutional neural network (CNN) denoisers \cite{Tirer2017ImageRB, Tirer2019SuperResolutionVI}. In \cite{Tirer2019BackProjectionBF}, the BP fidelity term is also mathematically analyzed and compared to LS for the case where $s(x)$ is the Tikhonov regularization (i.e. the $\ell_2$ prior).

Nowadays, CNNs are a very powerful tool for many computer vision tasks, including image restoration. For a given reconstruction task, CNNs can perform the inverse mapping from the observations to the signal domain and achieve state-of-the-art performances due to their ability to learn from large datasets. Researchers, motivated by deep learning great results, are applying deep neural networks to solve imaging inverse problems such as denoising \cite{Zhang2017BeyondAG,Mao2016ImageRU,Jain2008NaturalID,Xie2012ImageDA}, super-resolution \cite{Lim2017EnhancedDR,Dong2014ImageSU,Kim2015AccurateIS} and deblurring \cite{Sun2015LearningAC,Nah2016DeepMC,Chakrabarti2016ANA}.

However, the performance of deep learning methods tends to significantly drop once the observation model used in training mismatches the one in test time. This is the reason for the growing popularity of alternative methods, which are not biased to the observation model used in the offline training phase. One such example is the plug-and-play (P\&P) denoisers approach \cite{Venkatakrishnan2013PlugandPlayPF} and its successors, which have been proposed in the last few years. In the original P\&P paper \cite{Venkatakrishnan2013PlugandPlayPF}, the authors minimize \eqref{eq:cost_function} using an optimization algorithm that decouples the fidelity term and the prior term. % in the optimization procedure. 
%such that the prior can be handled 
They propose to avoid explicit formulation of $s(x)$, and instead, handle the prior 
by an arbitrary denoising operation (e.g. BM3D or CNN denoisers). %a pre-trained CNN denoiser). 
A related work is IDBP \cite{Tirer2017ImageRB}, which presents an alternative framework for solving inverse problems using off-the-shelf denoisers. This method requires less parameter tuning and implicitly uses the unique BP fidelity term \cite{Tirer2019BackProjectionBF}. Another recent variant of \cite{Venkatakrishnan2013PlugandPlayPF} %, named regularization by denoising \cite{Romano2016TheLE, bigdeli2017deep}, 
modifies the way that the denoising engine is used to regularize the inverse problem \cite{Romano2016TheLE, bigdeli2017deep}.

An additional disadvantage of deep learning originates from its requirement to often use a lot of data. Indeed, large datasets are used in the majority of works in this field. However, recently several deep learning methods, which are trained only using a single image, have been proposed and have demonstrated surprisingly good results. 
In Deep Image Prior (DIP) \cite{Ulyanov2017DeepIP} the authors use the deep CNN itself as a prior for various restoration tasks, e.g. denoising, super-resolution and inpainting. As part of this approach, a CNN is trained from scratch during test time, using only the degraded image, with no requirement for a training dataset.
Another work that trains a CNN super-resolver from scratch at test time is %Zero-Shot Super-Resolution (ZSSR) 
ZSSR \cite{Shocher2017ZeroShotSU}, which presents a combination of deep architectures with internal-example learning. Besides the test image, no other images are used---%, and all the training patches are taken from degraded pairs of the test image. 
all the pairs of training patches are extracted/synthesized from the test image. 
In SinGAN \cite{Shaham2019SinGANLA}, a generative model is learned from a single image. The SinGAN contains a pyramid of fully convolutional generative adversarial networks (GANs) \cite{goodfellow2014generative}, where each of them learns the patch distribution at different scales of the image. Both ZSSR and SinGAN focus on a specific task (super-resolution and samples generation, respectively), in contrast to the DIP approach. There are also works that incorporate training on external data with image-adaptation (via fine-tuning using the test image), such as IDBP-CNN-IA \cite{Tirer2019SuperResolutionVI} and IAGAN \cite{Hussein2019ImageAdaptiveGB}.

%{\bf Contribution.} 
In this paper, we focus on the DIP strategy. Most of the papers that apply this approach use the typical LS loss function and achieve rather limited performance and/or require a very large number of backprop iterations at test time \cite{Ulyanov2017DeepIP, van2018compressed, mataev2019deepred}. To mitigate these deficients, we propose to use a BP loss function instead of the LS loss. We demonstrate this approach for deblurring, using multiple kernels and noise levels, and present improved restoration results, which are reflected by higher PSNR and SSIM values and much better inference run-time. % (due to faster convergence).

\section{Background}

Before we turn to describe our method, we first describe in more details the DIP and BP strategies. 

\subsection{ Deep Image Prior (DIP)} 
% In image restoration tasks, two types of priors are mostly used: learned-prior and explicit-prior. The explicit-prior promotes images that are natural looking, although constraints such as “natural” are challenging to express mathematically. The learned-prior is often achieved by supervised training a deep convolution neural network 
In recent years, training deep convolution neural networks have become a common way to perform image restoration tasks. The popularity of this machine learning approach stems from the difficulty in accurately modelling natural images with explicit prior functions.
Typically, the implicit prior that is associated with a CNN is achieved by supervised learning using a large dataset, where the degraded images serve as the network’s inputs and the weights are optimized such that the  network’s outputs match the original images. 
%In \cite{Ulyanov2017DeepIP}, a combined prior is introduced, by constructing an explicit prior using a CNN. They study the prior implicitly captured by the choice of a particular network structure, before any of its parameters are learned. 

The Deep Image Prior (DIP) work \cite{Ulyanov2017DeepIP} has demonstrated a remarkable phenomenon: CNNs can be used for solving image restoration problems without any offline training and external data. % --- the network’s structure itself is sufficient for forcing a strong prior that allows to reconstruct an image from its degraded version. 
The DIP paper disputes the idea that supervised learning is mandatory for %building working image priors and 
restoring images with CNNs.
It shows that the network's architecture itself %captures much image statistics.
is sufficient for forcing a strong prior that allows reconstructing an image from its degraded version. %In fact, 
Therefore, 
%in order to perform restoration tasks, 
there is no need in large datasets and offline learning, as the image restoration can be performed only by using the single degraded image.

In DIP, the estimated image $x \in \Real^{3\times H\times W}$ is parameterized by
\begin{equation}
x = f_\theta(z),
\end{equation}
where $f_\theta(z)$ is (typically) a deep CNN with U-Net architecture \cite{Ronneberger2015UNetCN}, 
$z \in \Real^{C'\times H'\times W'}$ is a fixed tensor filled with uniform noise, % (though it can be perturbed randomly at every iteration (for some experiments)
and $\theta$ are the network parameters. 
Currently, DIP works consider a loss function that is given by the LS objective
\begin{equation}
\label{eq:loss_dip}
    \min_{\theta} \norm{ y - A f_\theta(z)}_2^2,
\end{equation}
and minimize it, with respect to $\theta$, using first-order methods such as SGD and Adam \cite{Kingma2014AdamAM}.
Note that overfitting $y$ may lead to $f_\theta(z)$ with artifacts. Therefore, in DIP the optimization process is terminated early.

It is interesting to note that the above strategy can be obtained from the general formulation in \eqref{eq:cost_function}, for the LS fidelity term
\begin{equation}
\label{eq:ls_fidelity}
    \ell(x,y) = \frac{1}{2} \norm{ y - A x }_2^2,
\end{equation}
and the following indicator prior
\begin{equation}
    s(x) = \begin{cases}
    0, & x = f_\theta(z)\\
    +\infty, & \text{otherwise}
    \end{cases}.
\end{equation}

% Note that the DIP approach can still be presented as a special case of \eqref{eq:cost_function},
% where 
% In DIP the neural network is interpreted as a parametrization
% \begin{equation}
% x = f_\theta(z)
% \end{equation}
% of an image $x \in \Real^{3\times H\times W}$. $z \in \Real^{C'\times H'\times W'}$ is a fixed tensor filled with uniform noise (though it can be perturbed randomly at every iteration (for some experiments), and $\theta$ are the network parameters. 
% Previous DIP works essentially use (\ref{eq:cost_function}) with LS as fidelity term and an indicator prior
% \begin{equation}
%     s(x) = \begin{cases}
%     0, & x = f_\theta\\
%     +\infty, & \text{otherwise}
%     \end{cases}
% \end{equation}

% In other words, the loss function of the neural network is given by

% \begin{equation}
% \label{eq:loss_dip}
%     \min_{\theta} \norm{ y - Af_\theta}_2^2,
% \end{equation}
% and is minimized (with respect to $\theta$) using first-order methods, such as SGD and Adam \cite{Kingma2014AdamAM}.

\subsection{The backprojection (BP) fidelity term}

Recently, the backprojection (BP) fidelity term has been proposed in \cite{Tirer2019BackProjectionBF} as an alternative to the widely used LS term \eqref{eq:ls_fidelity}. Under the practical assumptions that $m \le n$ and $\mathrm{rank}(A) = m$, the pseudoinverse of $A$ is given by $A^\dag = A^T (A A^T)^{-1}$, and the backprojection fidelity term is
\begin{equation}
\label{eq:bp_fidelity}
    \ell(x,y) = \frac{1}{2} \norm{A^\dag( y - Ax)}_2^2.
\end{equation}
This fidelity term encourages agreement between the projection of the optimization variable onto the row space of the linear operator (i.e. $A^\dag Ax$) and the pseudoinverse of the linear operator (”back-projection”) applied on the observations (i.e. $A^\dag y$). 
Note that \eqref{eq:bp_fidelity} can also be written as\footnote{The equivalence between \eqref{eq:bp_fidelity} and \eqref{eq:bp_fidelity2} can be observed by expanding the two quadratic forms.}
\begin{equation}
\label{eq:bp_fidelity2}
    \ell(x,y) = \frac{1}{2} \norm{(AA^T)^{-\frac{1}{2}}( y - Ax)}_2^2.
\end{equation}

It has been demonstrated in \cite{Tirer2019BackProjectionBF, Tirer2017ImageRB, Tirer2019SuperResolutionVI} that for different priors (e.g. TV, BM3D, and pre-trained CNNs) 
the BP fidelity term can yield better recoveries than LS for badly conditioned $A$ and requires fewer iterations of optimization algorithms (the improved convergence rate is also analyzed in \cite{tirer2020convergence}). 
Yet, when the singular values of $A$ are small, the performance advantage of BP is inversely proportional to the noise level.

\begin{figure*}
    \centering
    \begin{subfigure}[b]{0.32\textwidth}
        \centering
        \includegraphics[width=160pt]{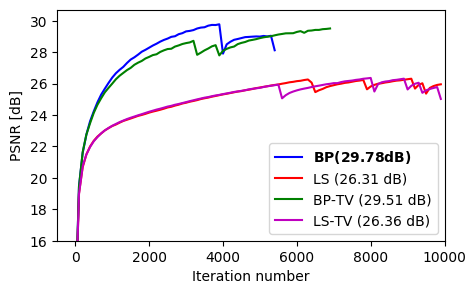}
        \caption{{\small Uniform kernel, $\sigma=\sqrt{0.3}$}}    
        \label{fig:uniform0.3}
    \end{subfigure}
    \begin{subfigure}[b]{0.32\textwidth}  
        \centering
        \includegraphics[width=160pt]{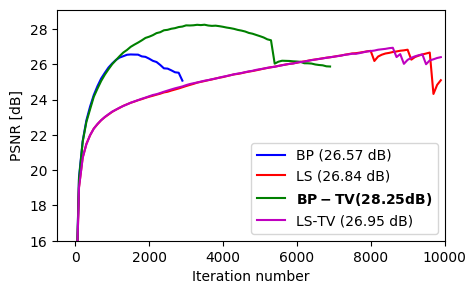}
        \caption{{\small Uniform kernel, $\sigma=\sqrt2$}} 
        \label{fig:uniform2}
    \end{subfigure}
    \begin{subfigure}[b]{0.32\textwidth}  
        \centering
        \includegraphics[width=160pt]{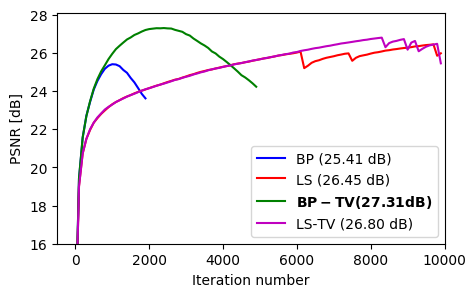}
        \caption{{\small Uniform kernel, $\sigma=\sqrt4$}}    
        \label{fig:uniform4}
    \end{subfigure}\\
    \begin{subfigure}[b]{0.32\textwidth}
        \centering
        \includegraphics[width=160pt]{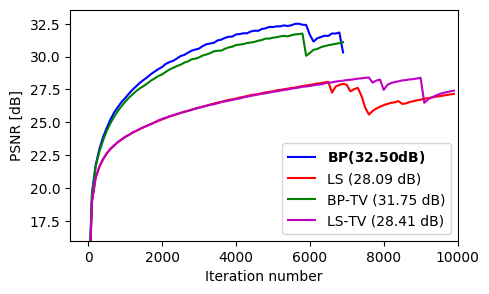}
        \caption{{\small Radial kernel, $\sigma=\sqrt{0.3}$}}    
        \label{fig:radial0.3}
    \end{subfigure}
    \begin{subfigure}[b]{0.32\textwidth}  
        \centering
        \includegraphics[width=160pt]{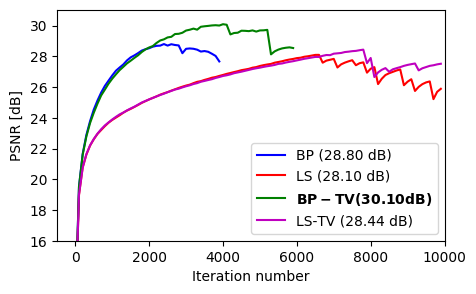}
        \caption{{\small Radial kernel, $\sigma=\sqrt2$}} 
        \label{fig:radial2}
    \end{subfigure}
    \begin{subfigure}[b]{0.32\textwidth}  
        \centering
        \includegraphics[width=160pt]{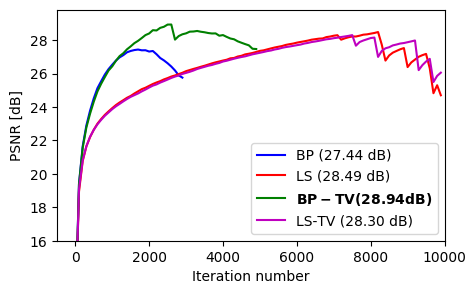}
        \caption{{\small Radial kernel, $\sigma=\sqrt4$}}    
        \label{fig:radial4}
    \end{subfigure}\\
    \begin{subfigure}[b]{0.32\textwidth}
        \centering
        \includegraphics[width=160pt]{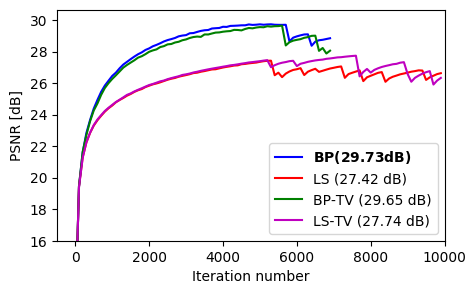}
        \caption{{\small Gaussian kernel, $\sigma=\sqrt{0.3}$}}    
        \label{fig:gaus0.3}
    \end{subfigure}
    \begin{subfigure}[b]{0.32\textwidth}  
        \centering
        \includegraphics[width=160pt]{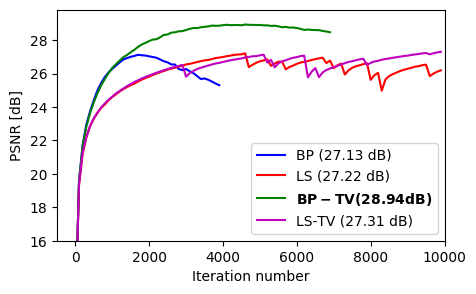}
        \caption{{\small Gaussian kernel, $\sigma=\sqrt2$}} 
        \label{fig:gaus2}
    \end{subfigure}
    %\hfill
    \begin{subfigure}[b]{0.32\textwidth}  
        \centering
        \includegraphics[width=160pt]{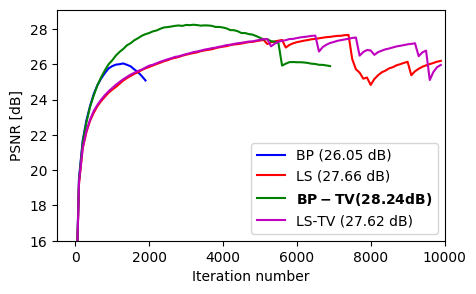}
        \caption{{\small Gaussian kernel, $\sigma=\sqrt4$}}    
        \label{fig:gaus4}
    \end{subfigure}\\
     \caption{Deblurring results (averaged over Set14) for: Uniform kernel in (a)-(c), Radial kernel in (d)-(f) and Gaussian kernel in (g)-(i), for noise levels $\sigma=\sqrt{0.3},\sqrt2$ and $\sqrt4$. Note that BP achieves best PSNR in all kernels with $\sigma=\sqrt{0.3}$, and BP-TV achieves best PSNR in all kernels with higher noise levels ($\sigma=\sqrt2,\sqrt4$).}
    \label{fig:comparison_3kernels_3sigma}
\end{figure*}

\section{Back-Projection based Deep Image Prior}

In this paper, we demonstrate that using the BP fidelity term improves the performance of standard DIP, which uses the LS fidelity term as the loss function. The use of BP fidelity term \eqref{eq:bp_fidelity2} in DIP leads to the following cost function
% \begin{equation}
% \label{eq:loss_bp1}
%     \min_{\theta} \norm{A^\dag( y - Af_\theta)}_2^2.
% \end{equation}
% Note that (\ref{eq:loss_bp1}) can also be written as\footnote{The equivalence between (\ref{eq:loss_bp1}) and (\ref{eq:loss_bp2}) can be observed by expanding the two quadratic forms.}
\begin{equation}
\label{eq:loss_bp2}
    \min_{\theta} \norm{(AA^T)^{-\frac{1}{2}}( y - A f_\theta(z) )}_2^2.
\end{equation}

For the image deblurring task, $A$ represents convolution with a blur kernel $h$. %, and can be fastly computed via Fast Fourier Transform (FFT).
Therefore, in this case, $AA^T$ represents convolution with the filter $h*flip(h)$. This operator, as well as its square root inverse $(AA^T)^{-\frac{1}{2}}$, has a very fast implementation using the Fast Fourier Transform (FFT) \cite{cooley1965algorithm}. % be implemented by FFT. % as well.
To conclude, the loss function \eqref{eq:loss_bp2} can be 
% computing the loss function \eqref{eq:loss_bp2} (and its gradients) can be efficiently done using Fast Fourier Transform (FFT). In this case, $AA^T$ represents filtering with $h*flip(h)$, and therefore $(AA^T)^{-\frac{1}{2}}$ can be implemented by FFT as well. 
%To conclude, 
efficiently implemented by\footnote{Note that this formulation also allows to easily obtain the loss gradients using popular software packages, such as TensorFlow and PyTorch.} % (\ref{eq:loss_bp2}) by 
\begin{equation}
\label{eq:loss_bp-dip_fft}
    \min_{\theta} \norm{\mathcal{F}^*(\frac{1}{\sqrt{|\mathcal{F}(h)|^2+\epsilon_1\sigma^2+\epsilon_2}}\mathcal{F}( y - h*f_\theta))}_2^2,
\end{equation}
where $h$ is a blur kernel, $\sigma$ is the noise level in \eqref{eq:inverse_prob}, and $\mathcal{F}$, $\mathcal{F}^*$ stand for Fourier transform and inverse Fourier transform, respectively. $\epsilon_1$ and $\epsilon_2$ are regularization parameters, which are required since $A$ is ill-conditioned in case of deblurring, and the scenarios include noise. 
Note that FFT implementation of the operator $AA^T$ and its inverse can be done in more restoration tasks, such as in super-resolution (e.g. see \cite{shady2019correction}).

\begin{figure*}
    \center
    \begin{subfigure}[h]{0.32\columnwidth}
        \centering
        \includegraphics[trim={110pt 0pt 0pt 230pt}, clip, width=2.9cm, height=2.9cm]{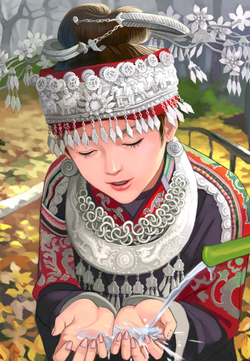}
        \caption{Ground truth}
    \end{subfigure}
    \begin{subfigure}[h]{0.32\columnwidth}
        \centering
        \includegraphics[trim={110pt 0pt 0pt 230pt}, clip, width=2.9cm, height=2.9cm]{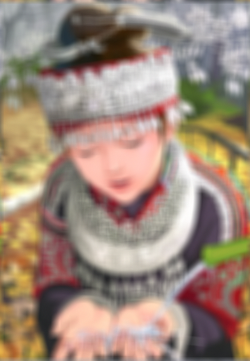}
        \caption{Blurred image}
    \end{subfigure}
    \begin{subfigure}[h]{0.32\columnwidth}
        \centering
        \includegraphics[trim={110pt 0pt 0pt 230pt}, clip, width=2.9cm, height=2.9cm]{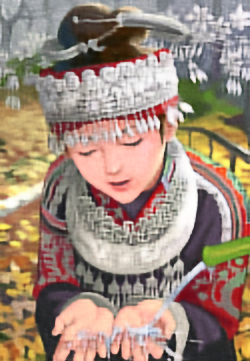}
        \caption{LS}
    \end{subfigure}
    \begin{subfigure}[h]{0.32\columnwidth}
        \centering
        \includegraphics[trim={110pt 0pt 0pt 230pt}, clip, width=2.9cm, height=2.9cm]{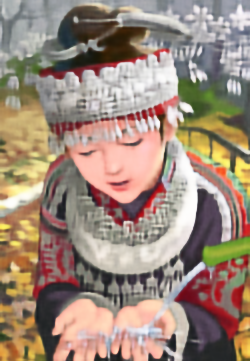}
        \caption{LS-TV}
    \end{subfigure}
    \begin{subfigure}[h]{0.32\columnwidth}
        \centering
        \includegraphics[trim={110pt 0pt 0pt 230pt}, clip, width=2.9cm, height=2.9cm]{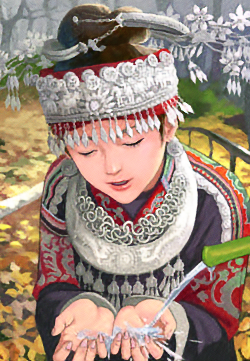}
        \caption{BP}
    \end{subfigure}
    \begin{subfigure}[h]{0.32\columnwidth}
        \centering
        \includegraphics[trim={110pt 0pt 0pt 230pt}, clip, width=2.9cm, height=2.9cm]{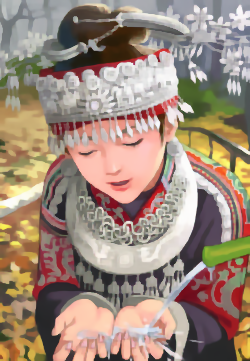}
        \caption{BP-TV}
    \end{subfigure}
    \caption{Deblurring using BP-DIP and LS-DIP (with and without TV). Uniform kernel, $\sigma=\sqrt{0.3}$}
    \label{fig:uniform_0.3}
\end{figure*}

\begin{figure*}
    \center
    \begin{subfigure}[h]{0.32\columnwidth}
        \centering
        \includegraphics[trim={400pt 300pt 100pt 50pt}, clip, width=2.9cm, height=2.9cm]{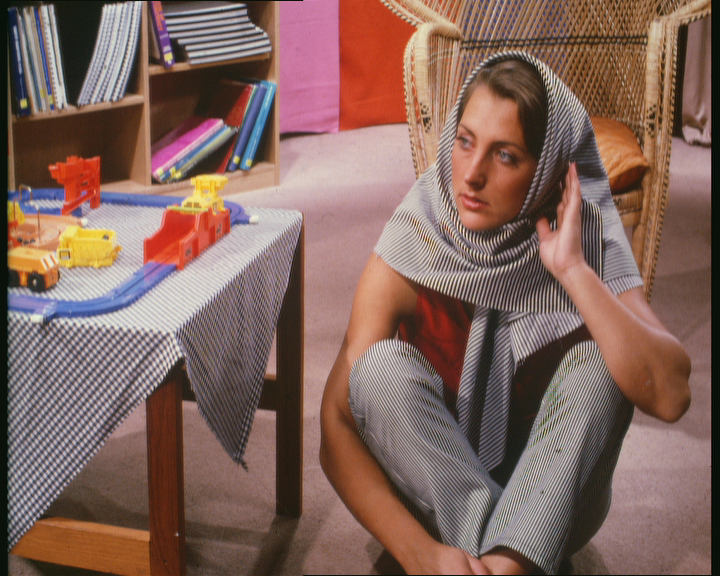}
        \caption{Ground truth}
    \end{subfigure}
    \begin{subfigure}[h]{0.32\columnwidth}
        \centering
        \includegraphics[trim={400pt 300pt 100pt 50pt}, clip, width=2.9cm, height=2.9cm]{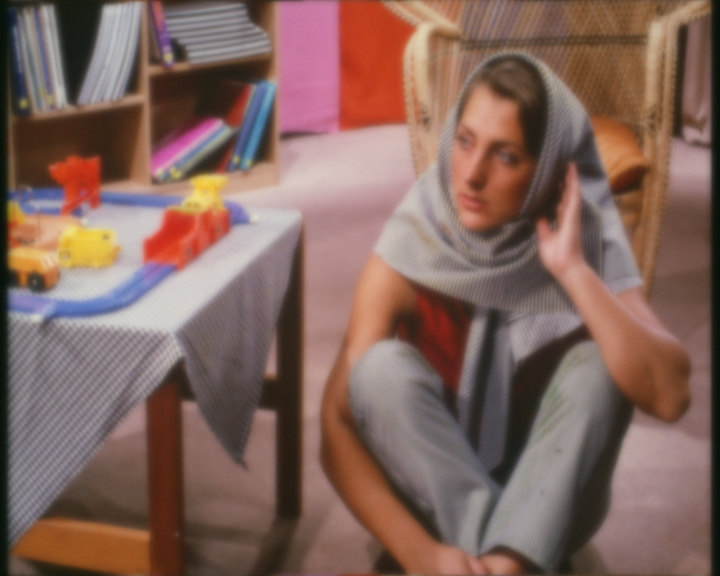}
        \caption{Blurred image}
    \end{subfigure}
    \begin{subfigure}[h]{0.32\columnwidth}
        \centering
        \includegraphics[trim={400pt 300pt 100pt 50pt}, clip, width=2.9cm, height=2.9cm]{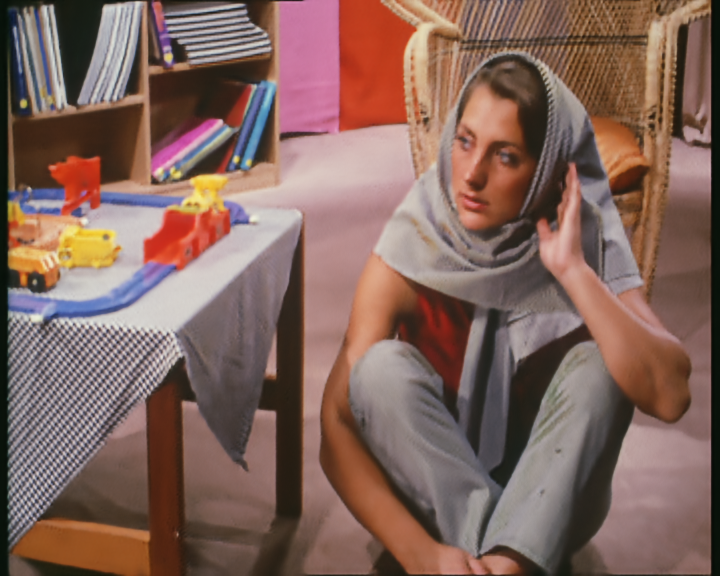}
        \caption{LS}
    \end{subfigure}
    \begin{subfigure}[h]{0.32\columnwidth}
        \centering
        \includegraphics[trim={400pt 300pt 100pt 50pt}, clip, width=2.9cm, height=2.9cm]{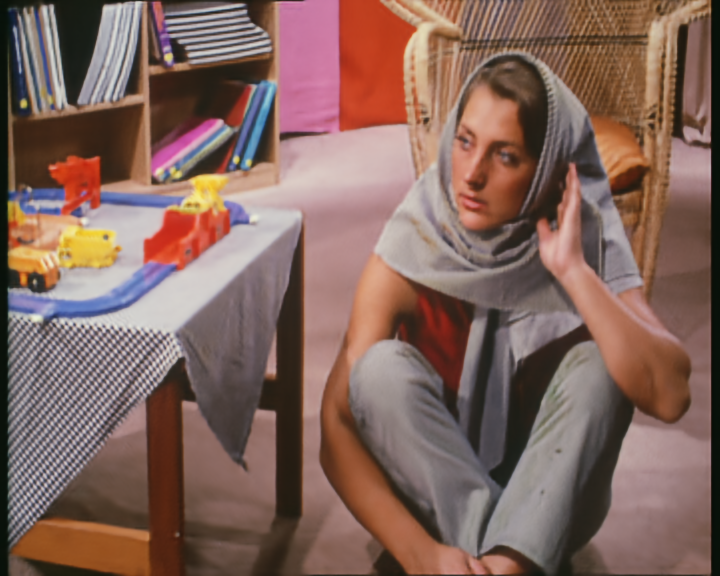}
        \caption{LS-TV}
    \end{subfigure}
    \begin{subfigure}[h]{0.32\columnwidth}
        \centering
        \includegraphics[trim={400pt 300pt 100pt 50pt}, clip, width=2.9cm, height=2.9cm]{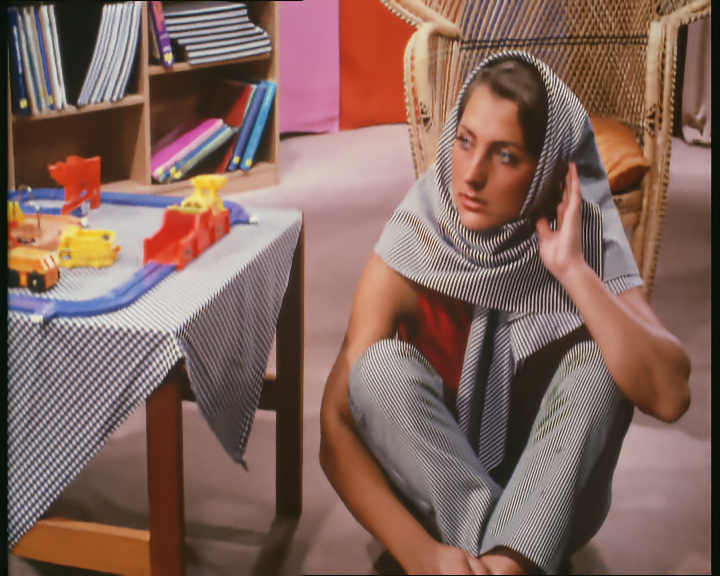}
        \caption{BP}
    \end{subfigure}
    \begin{subfigure}[h]{0.32\columnwidth}
        \centering
        \includegraphics[trim={400pt 300pt 100pt 50pt}, clip, width=2.9cm, height=2.9cm]{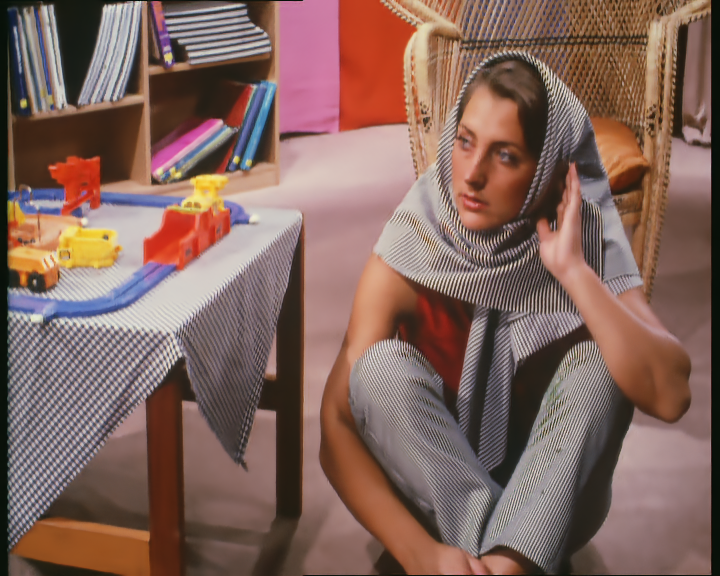}
        \caption{BP-TV}
    \end{subfigure}
    \caption{Deblurring using BP-DIP and LS-DIP (with and without TV). Radial kernel, $\sigma=\sqrt{2}$}
    \label{fig:radial_2}
\end{figure*}

\begin{figure*}
    \center
    \begin{subfigure}[h]{0.32\columnwidth}
        \centering
        \includegraphics[trim={30pt 200pt 250pt 240pt}, clip, width=2.9cm, height=2.9cm]{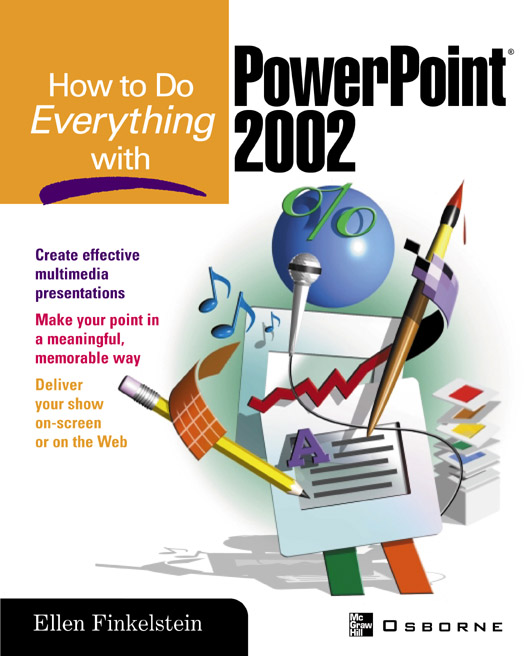}
        \caption{Ground truth}
    \end{subfigure}
    \begin{subfigure}[h]{0.32\columnwidth}
        \centering
        \includegraphics[trim={30pt 200pt 250pt 240pt}, clip, width=2.9cm, height=2.9cm]{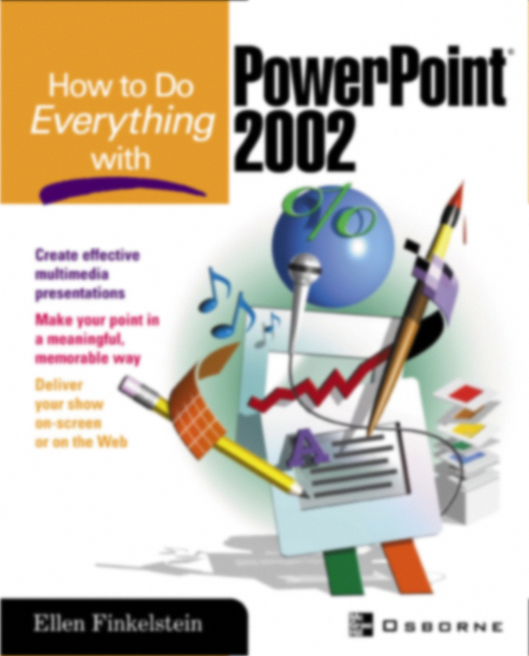}
        \caption{Blurred image}
    \end{subfigure}
    \begin{subfigure}[h]{0.32\columnwidth}
        \centering
        \includegraphics[trim={30pt 200pt 250pt 240pt}, clip, width=2.9cm, height=2.9cm]{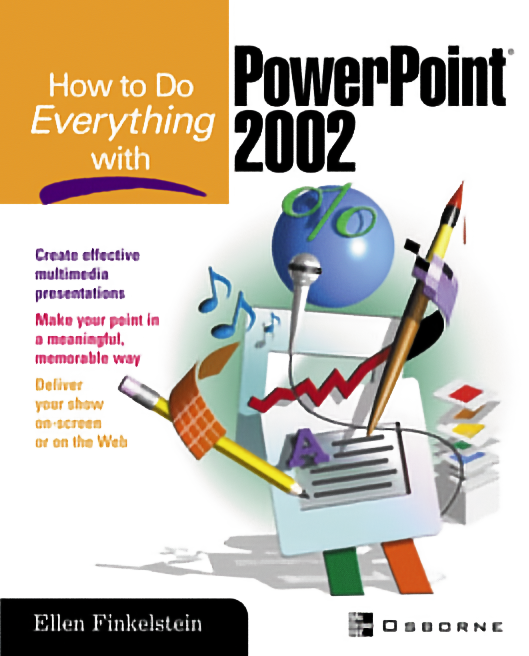}
        \caption{LS}
    \end{subfigure}
    \begin{subfigure}[h]{0.32\columnwidth}
        \centering
        \includegraphics[trim={30pt 200pt 250pt 240pt}, clip, width=2.9cm, height=2.9cm]{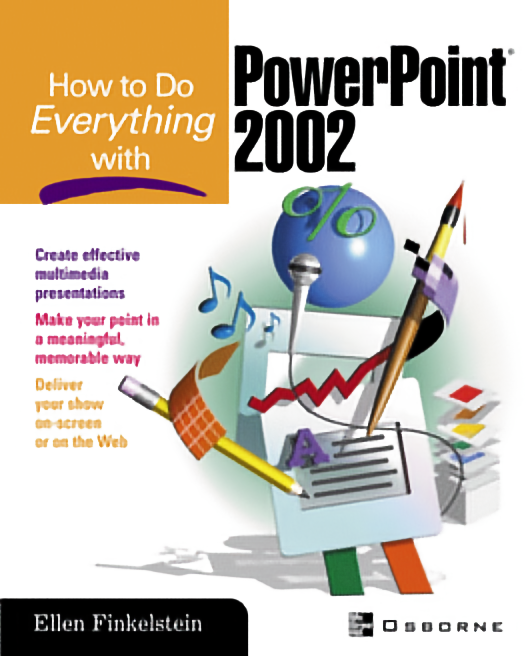}
        \caption{LS-TV}
    \end{subfigure}
    \begin{subfigure}[h]{0.32\columnwidth}
        \centering
        \includegraphics[trim={30pt 200pt 250pt 240pt}, clip, width=2.9cm, height=2.9cm]{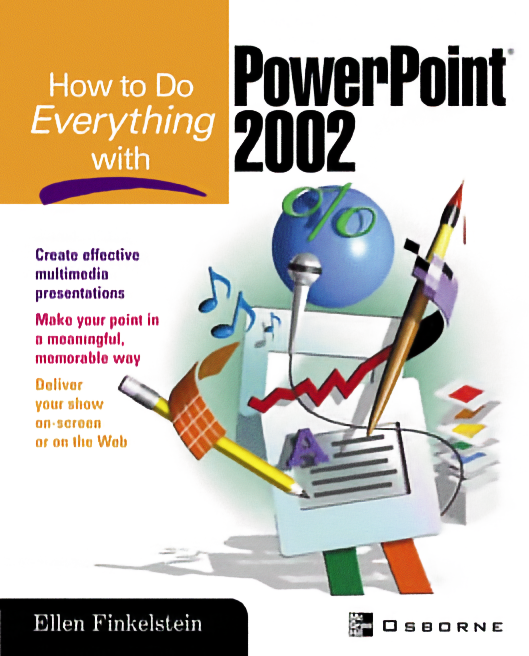}
        \caption{BP}
    \end{subfigure}
    \begin{subfigure}[h]{0.32\columnwidth}
        \centering
        \includegraphics[trim={30pt 200pt 250pt 240pt}, clip, width=2.9cm, height=2.9cm]{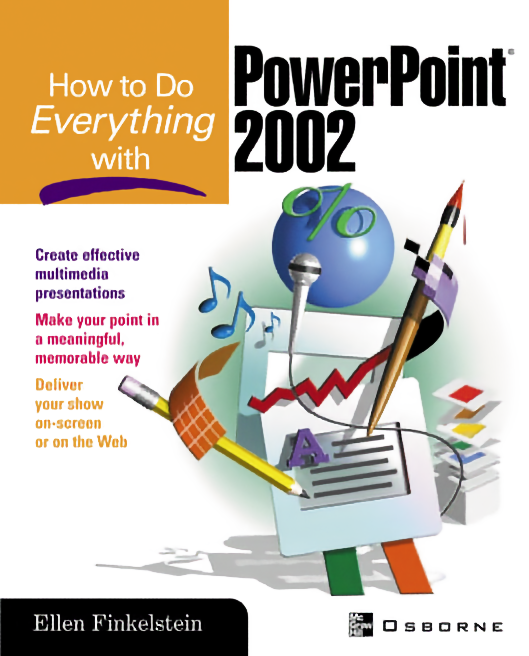}
        \caption{BP-TV}
    \end{subfigure}
    \caption{Deblurring using BP-DIP and LS-DIP (with and without TV). Gaussian kernel, $\sigma=\sqrt{2}$}
    \label{fig:gaus_2}
\end{figure*}

\begin{figure*}
    \center
    \begin{subfigure}[h]{0.32\columnwidth}
        \centering
        \includegraphics[trim={150pt 160pt 150pt 0pt}, clip, width=2.9cm, height=2.9cm]{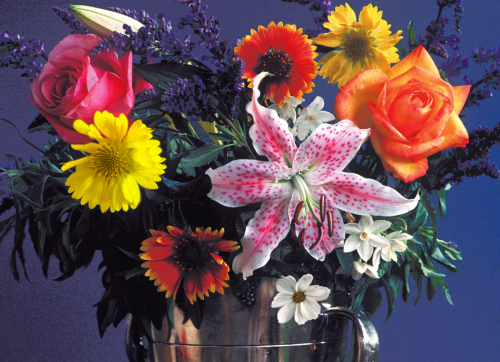}
        \caption{Ground truth}
    \end{subfigure}
    \begin{subfigure}[h]{0.32\columnwidth}
        \centering
        \includegraphics[trim={150pt 160pt 150pt 0pt}, clip, width=2.9cm, height=2.9cm]{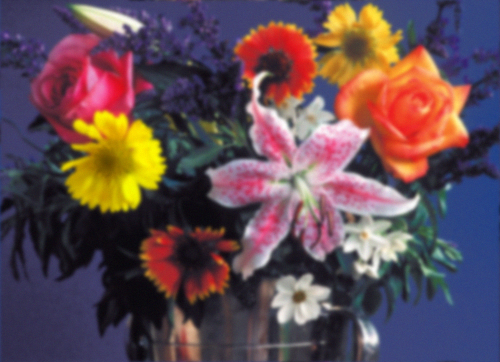}
        \caption{Blurred image}
    \end{subfigure}
    \begin{subfigure}[h]{0.32\columnwidth}
        \centering
        \includegraphics[trim={150pt 160pt 150pt 0pt}, clip, width=2.9cm, height=2.9cm]{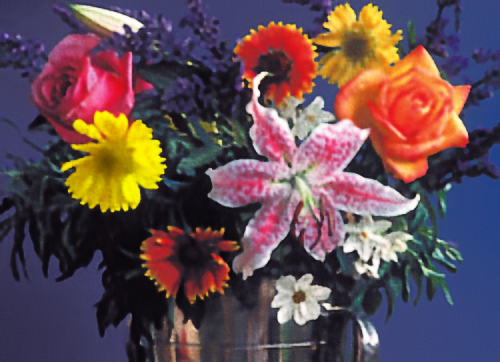}
        \caption{LS}
    \end{subfigure}
    \begin{subfigure}[h]{0.32\columnwidth}
        \centering
        \includegraphics[trim={150pt 160pt 150pt 0pt}, clip, width=2.9cm, height=2.9cm]{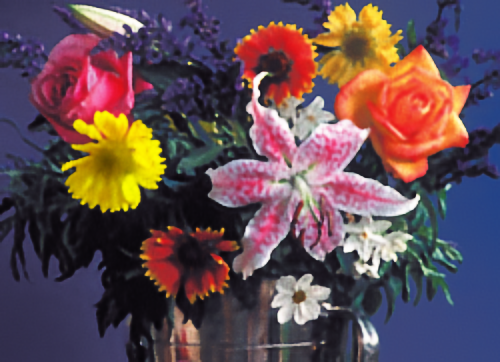}
        \caption{LS-TV}
    \end{subfigure}
    \begin{subfigure}[h]{0.32\columnwidth}
        \centering
        \includegraphics[trim={150pt 160pt 150pt 0pt}, clip, width=2.9cm, height=2.9cm]{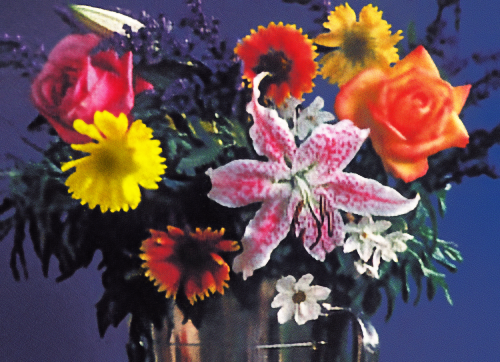}
        \caption{BP}
    \end{subfigure}
    \begin{subfigure}[h]{0.32\columnwidth}
        \centering
        \includegraphics[trim={150pt 160pt 150pt 0pt}, clip, width=2.9cm, height=2.9cm]{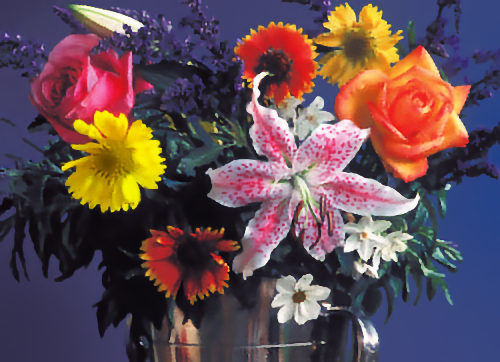}
        \caption{BP-TV}
    \end{subfigure}
    \caption{Deblurring using BP-DIP and LS-DIP (with and without TV). Gaussian kernel, $\sigma=\sqrt{4}$}
    \label{fig:gaus_4}
\end{figure*}

As explained in \cite{Tirer2019BackProjectionBF}, in scenarios such as deblurring, where $A$ has many singular values that are much smaller than $1$, the BP term is more sensitive to noise than the LS term. Therefore, the inversion of $AA^T$ has to be regularized. The higher the noise level in $y$ is, the more sensitive the expression $\frac{1}{\sqrt{|\mathcal{F}(h)|^2+\epsilon_1\sigma^2+\epsilon_2}}\mathcal{F}(y)$ in (\ref{eq:loss_bp-dip_fft}). %Therefore, despite the advantages of BP over LS for deblurring \cite{Tirer2019BackProjectionBF}, at some noise level using LS is likely to be preferable. 
% While for many priors this "phase transition" happens at quite large noise levels \cite{Tirer2019BackProjectionBF, Tirer2017ImageRB, Tirer2019SuperResolutionVI}, % (e.g. see deblurring Scenario 4 in \cite{Tirer2017ImageRB}), 
% our experiments suggest that for DIP it happens at a rather small noise level.

\begin{table*}[t]
  \begin{center}
  \caption{Deblurring results (PSNR [dB] / SSIM averaged over Set14) of the different methods}
    \label{tab:table1}
    \begin{tabular}{|c|c|c|c|c|c|c|c|c|c|}
    \hline
    ~ & \multicolumn{3}{c|}{Uniform kernel} & \multicolumn{3}{c|}{Radial kernel} & \multicolumn{3}{c|}{Gaussian kernel}\\
    %\textbf{LS} & \textbf{LS-TV} & \textbf{BP} & \textbf{BP-DIP} \\
    \hline
    ~ & $\sigma=\sqrt{0.3}$ &  $\sigma=\sqrt{2}$ &  $\sigma=\sqrt{4}$ & $\sigma=\sqrt{0.3}$ &  $\sigma=\sqrt{2}$ &  $\sigma=\sqrt{4}$ & $\sigma=\sqrt{0.3}$ &  $\sigma=\sqrt{2}$ &  $\sigma=\sqrt{4}$ \\
    \hline
    LS & 26.31 / 0.83 & 26.84 / 0.84 & 26.45 / 0.83 & 28.09 / 0.88 & 28.10 / 0.88 & 28.49 / 0.87 & 27.42 / 0.87 & 27.22 / 0.86 & 27.66 / 0.86 \\ 
    \hline
   LS-TV & 26.36 / 0.83 & 26.95 / 0.84 & 26.80 / 0.84 & 28.41 / 0.88 & 28.44 / 0.88 & 28.30 / 0.88 & 27.74 / 0.86 & 27.31 / 0.85 & 27.62 / 0.86 \\ 
    \hline
   BP & \textbf{29.78 / 0.91} & 26.57 / 0.84 & 25.41 / 0.80 & \textbf{32.50 / 0.95} & 28.80 / 0.89 & 27.44 / 0.86 & \textbf{29.73 / 0.91} & 27.13 / 0.85 & 26.05 / 0.82 \\ 
    \hline
    BP-TV & 29.51 / 0.90 & \textbf{28.25 / 0.88} & \textbf{27.31 / 0.86} & 31.75 / 0.93 & \textbf{30.10 / 0.91} & \textbf{28.94 / 0.89} & 29.65 / 0.91 & \textbf{28.94 / 0.90} & \textbf{28.24 / 0.88} \\ 
    \hline
    \end{tabular}
  \end{center}
\end{table*}

In order to mitigate the sensitivity to noise of BP for DIP (which seems somewhat increased compared to other priors \cite{Tirer2019BackProjectionBF, Tirer2017ImageRB, Tirer2019SuperResolutionVI}), we propose to add TV regularization \cite{Rudin1992NonlinearTV} to the loss function \eqref{eq:loss_bp-dip_fft}. The TV term encourages piecewise smoothness in the image and has led to a more balanced restoration in our experiments. The (anisotropic) TV loss, used in this work, is given by
\begin{equation}
\label{eq:tv}
     \sum_{i,j} |x_{{i+1},j}-x_{i,j}|+|x_{i,{j+1}}-x_{i,j}|
\end{equation}
for a two-dimensional signal $x$.

\section{Experiments}

We demonstrate our approach on the deblurring task, which aims at recovering the original, sharp image from a blurred image. The performance of the deblurring process is compared between two cases: DIP with backprojection loss and DIP with least squares loss, denoted by BP-DIP and LS-DIP, respectively. We use 3 different blur kernels; radial, Gaussian and uniform, with 3 different noise levels: $\sqrt{0.3}, \sqrt2$ and $\sqrt4$. 
The radial kernel is of size 15 × 15 and can be written as $\frac{1}{1+x_1^2+x_2^2}, x_1,x_2=-7,...,7$, the Gaussian kernel is of size 15 × 15 with STD 1.6, and the uniform kernel is of size 9 x 9.
All three kernels are normalized to the sum of 1.

The hyper-parameters and CNN are chosen to be the same as in the original DIP paper \cite{Ulyanov2017DeepIP}, such that all of the experiments are performed using a U-Net architecture \cite{Ronneberger2015UNetCN} with skip-connections where the input and the output have the same spatial size, the optimiser is Adam \cite{Kingma2014AdamAM} and the learning rate is 0.01. The deblurring performance is evaluated on Set14 dataset. 

The weight of the TV regularizer was chosen as the best value for BP and LS separately: $1e-3$ for BP (all kernels), $1e-5$ and $1e-6$ for LS with Gaussian/radial kernels and uniform kernel, respectively. The $\epsilon_1$ and $\epsilon_2$ values in equation (\ref{eq:loss_bp-dip_fft}) were chosen as $0.01$ and $1e-3$ respectively.

Curves of PNSR vs. iteration number for all 9 experiments are displayed in Figure \ref{fig:comparison_3kernels_3sigma}, and the final PNSR/SSIM are also displayed in Table~\ref{tab:table1}. 
Several visual examples are presented in Figures \ref{fig:uniform_0.3}-\ref{fig:gaus_4}. 
It is apparent that: (a) with BP loss, DIP yields higher PSNR than with LS loss; % (the same holds for SSIM);
(b) BP-DIP %loss converges 
reaches its peak PSNR 
faster than LS-DIP; (c) In most cases, adding the TV term improves the PSNR for both LS and BP, however, the bigger improvement is seen with BP-DIP; (d) As in the original DIP paper, early stopping is needed, otherwise the images are corrupted with noise. 

Notice that when $\sigma=\sqrt{0.3}$, BP presents the best PSNR out of the 4 methods (BP, BP-TV, LS, LS-TV) and there is, in fact, no need in adding the TV term. However, when $\sigma$ is higher, BP-TV presents the best PSNR out of the 4 methods. In general, when the noise level rises, the gap between BP and LS results reduces and adding TV to the loss term usually boosts the results. Also notice that the PSNR of LS-DIP starts descending at some iteration, similarly to the PSNR of BP-DIP (i.e. both of them require early stopping). However, for LS-DIP it happens at a much later iteration. This behavior shows than when the number of iterations is tuned for best performance, BP-DIP (even with TV) is faster than the LS-DIP.

% Notice that in a few cases the LS and LS-TV graphs continue to ascend after 7000 iterations. We conducted two extra experiments with 10000 iterations, for LS uniform kernel with $\sigma=\sqrt{0.3}$ and LS-TV radial kernel with $\sigma=\sqrt{2}$, and saw that maximum value in both cases appeared after $\sim8200$ iterations with $\sim0.5dB$ improvement. Meaning that even when running extra iterations for LS and LS-TV, still BP demonstrates the best results.

\section{Conclusion}

In this work, we examined the influence of the backprojection (BP) fidelity term on Deep Image Prior (DIP). We conducted multiple deblurring experiments, using various blur kernels and noise levels and achieved significant improvement over the DIP work, both in PSNR and in the required number of optimization iterations (and thus in the inference run-time). 
Our approach presents another empirical evidence that untrained CNNs can reconstruct a clean and sharp image using only its degraded version. %from the degraded one using a single image, 
%and can demonstrate
Yet, it demonstrates that 
very good results can be obtained even after a relatively small number of iterations. Future work includes examining the same concept with other restoration tasks, e.g. super-resolution.

% \section*{Acknowledgment}
% This work was supported by...

\bibliographystyle{IEEEtran}%{plain}
{\bibliography{mybib}}

\begin{thebibliography}{10}
\providecommand{\url}[1]{#1}
\csname url@samestyle\endcsname
\providecommand{\newblock}{\relax}
\providecommand{\bibinfo}[2]{#2}
\providecommand{\BIBentrySTDinterwordspacing}{\spaceskip=0pt\relax}
\providecommand{\BIBentryALTinterwordstretchfactor}{4}
\providecommand{\BIBentryALTinterwordspacing}{\spaceskip=\fontdimen2\font plus
\BIBentryALTinterwordstretchfactor\fontdimen3\font minus
  \fontdimen4\font\relax}
\providecommand{\BIBforeignlanguage}[2]{{%
\expandafter\ifx\csname l@#1\endcsname\relax
\typeout{** WARNING: IEEEtran.bst: No hyphenation pattern has been}%
\typeout{** loaded for the language `#1'. Using the pattern for}%
\typeout{** the default language instead.}%
\else
\language=\csname l@#1\endcsname
\fi
#2}}
\providecommand{\BIBdecl}{\relax}
\BIBdecl

\bibitem{Rudin1992NonlinearTV}
L.~I. Rudin, S.~Osher, and E.~Fatemi, ``Nonlinear total variation based noise
  removal algorithms,'' 1992.

\bibitem{Dabov2007ImageDB}
K.~Dabov, A.~Foi, V.~Katkovnik, and K.~O. Egiazarian, ``Image denoising by
  sparse 3-d transform-domain collaborative filtering,'' \emph{IEEE
  Transactions on Image Processing}, vol.~16, pp. 2080--2095, 2007.

\bibitem{Bora2017CompressedSU}
A.~Bora, A.~Jalal, E.~Price, and A.~G. Dimakis, ``Compressed sensing using
  generative models,'' in \emph{ICML}, 2017.

\bibitem{Buades2005ANA}
A.~Buades, B.~Coll, and J.-M. Morel, ``A non-local algorithm for image
  denoising,'' \emph{2005 IEEE Computer Society Conference on Computer Vision
  and Pattern Recognition}, vol.~2, pp. 60--65 vol. 2, 2005.

\bibitem{Glasner2009SuperresolutionFA}
D.~Glasner, S.~Bagon, and M.~Irani, ``Super-resolution from a single image,''
  \emph{2009 IEEE 12th International Conference on Computer Vision}, pp.
  349--356, 2009.

\bibitem{Tirer2019BackProjectionBF}
T.~Tirer and R.~Giryes, ``Back-projection based fidelity term for ill-posed
  linear inverse problems,'' \emph{IEEE Transactions on Image Processing},
  vol.~29, no.~1, pp. 6164--6179, 2020.

\bibitem{Tirer2017ImageRB}
------, ``Image restoration by iterative denoising and backward projections,''
  \emph{IEEE Transactions on Image Processing}, vol.~28, pp. 1220--1234, 2017.

\bibitem{Tirer2019SuperResolutionVI}
------, ``Super-resolution via image-adapted denoising {CNNs}: Incorporating
  external and internal learning,'' \emph{IEEE Signal Processing Letters},
  vol.~26, pp. 1080--1084, 2019.

\bibitem{Zhang2017BeyondAG}
K.~Zhang, W.~Zuo, Y.~Chen, D.~Meng, and L.~Zhang, ``Beyond a gaussian denoiser:
  Residual learning of deep cnn for image denoising,'' \emph{IEEE Transactions
  on Image Processing}, vol.~26, pp. 3142--3155, 2017.

\bibitem{Mao2016ImageRU}
X.-J. Mao, C.~Shen, and Y.-B. Yang, ``Image restoration using very deep
  convolutional encoder-decoder networks with symmetric skip connections,'' in
  \emph{NIPS}, 2016.

\bibitem{Jain2008NaturalID}
V.~Jain and H.~S. Seung, ``Natural image denoising with convolutional
  networks,'' in \emph{NIPS}, 2008.

\bibitem{Xie2012ImageDA}
J.~Xie, L.~Xu, and E.~Chen, ``Image denoising and inpainting with deep neural
  networks,'' in \emph{NIPS}, 2012.

\bibitem{Lim2017EnhancedDR}
B.~Lim, S.~Son, H.~Kim, S.~Nah, and K.~M. Lee, ``Enhanced deep residual
  networks for single image super-resolution,'' \emph{2017 IEEE Conference on
  Computer Vision and Pattern Recognition Workshops (CVPRW)}, pp. 1132--1140,
  2017.

\bibitem{Dong2014ImageSU}
C.~Dong, C.~C. Loy, K.~He, and X.~Tang, ``Image super-resolution using deep
  convolutional networks,'' \emph{IEEE Transactions on Pattern Analysis and
  Machine Intelligence}, vol.~38, pp. 295--307, 2014.

\bibitem{Kim2015AccurateIS}
J.~Kim, J.~K. Lee, and K.~M. Lee, ``Accurate image super-resolution using very
  deep convolutional networks,'' \emph{2016 IEEE Conference on Computer Vision
  and Pattern Recognition (CVPR)}, pp. 1646--1654, 2015.

\bibitem{Sun2015LearningAC}
J.~Sun, W.~Cao, Z.~Xu, and J.~Ponce, ``Learning a convolutional neural network
  for non-uniform motion blur removal,'' \emph{2015 IEEE Conference on Computer
  Vision and Pattern Recognition (CVPR)}, pp. 769--777, 2015.

\bibitem{Nah2016DeepMC}
S.~Nah, T.~H. Kim, and K.~M. Lee, ``Deep multi-scale convolutional neural
  network for dynamic scene deblurring,'' \emph{2017 IEEE Conference on
  Computer Vision and Pattern Recognition (CVPR)}, pp. 257--265, 2016.

\bibitem{Chakrabarti2016ANA}
A.~Chakrabarti, ``A neural approach to blind motion deblurring,'' \emph{ArXiv},
  vol. abs/1603.04771, 2016.

\bibitem{Venkatakrishnan2013PlugandPlayPF}
S.~V. Venkatakrishnan, C.~A. Bouman, and B.~Wohlberg, ``Plug-and-play priors
  for model based reconstruction,'' \emph{2013 IEEE Global Conference on Signal
  and Information Processing}, pp. 945--948, 2013.

\bibitem{Romano2016TheLE}
Y.~Romano, M.~Elad, and P.~Milanfar, ``The little engine that could:
  Regularization by denoising (red),'' \emph{ArXiv}, vol. abs/1611.02862, 2016.

\bibitem{bigdeli2017deep}
S.~A. Bigdeli, M.~Zwicker, P.~Favaro, and M.~Jin, ``Deep mean-shift priors for
  image restoration,'' in \emph{Advances in Neural Information Processing
  Systems}, 2017, pp. 763--772.

\bibitem{Ulyanov2017DeepIP}
D.~Ulyanov, A.~Vedaldi, and V.~S. Lempitsky, ``Deep image prior,'' \emph{2018
  IEEE/CVF Conference on Computer Vision and Pattern Recognition}, pp.
  9446--9454, 2017.

\bibitem{Shocher2017ZeroShotSU}
A.~Shocher, N.~Cohen, and M.~Irani, ``Zero-shot super-resolution using deep
  internal learning,'' \emph{2018 IEEE/CVF Conference on Computer Vision and
  Pattern Recognition}, pp. 3118--3126, 2017.

\bibitem{Shaham2019SinGANLA}
T.~R. Shaham, T.~Dekel, and T.~Michaeli, ``Singan: Learning a generative model
  from a single natural image,'' \emph{ArXiv}, vol. abs/1905.01164, 2019.

\bibitem{goodfellow2014generative}
I.~Goodfellow, J.~Pouget-Abadie, M.~Mirza, B.~Xu, D.~Warde-Farley, S.~Ozair,
  A.~Courville, and Y.~Bengio, ``Generative adversarial nets,'' in \emph{NIPS},
  2014, pp. 2672--2680.

\bibitem{Hussein2019ImageAdaptiveGB}
S.~{Abu Hussein}, T.~Tirer, and R.~Giryes, ``Image-adaptive {GAN} based
  reconstruction,'' \emph{AAAI Conference on Artificial Intelligence}, 2020.

\bibitem{van2018compressed}
D.~Van~Veen, A.~Jalal, M.~Soltanolkotabi, E.~Price, S.~Vishwanath, and A.~G.
  Dimakis, ``Compressed sensing with deep image prior and learned
  regularization,'' \emph{arXiv preprint arXiv:1806.06438}, 2018.

\bibitem{mataev2019deepred}
G.~Mataev, P.~Milanfar, and M.~Elad, ``Deepred: Deep image prior powered by
  red,'' in \emph{Proceedings of the IEEE International Conference on Computer
  Vision Workshops}, 2019.

\bibitem{Ronneberger2015UNetCN}
O.~Ronneberger, P.~Fischer, and T.~Brox, ``U-net: Convolutional networks for
  biomedical image segmentation,'' \emph{ArXiv}, vol. abs/1505.04597, 2015.

\bibitem{Kingma2014AdamAM}
D.~P. Kingma and J.~Ba, ``Adam: A method for stochastic optimization,''
  \emph{CoRR}, vol. abs/1412.6980, 2014.

\bibitem{tirer2020convergence}
T.~Tirer and R.~Giryes, ``On the convergence rate of projected gradient descent
  for a back-projection based objective,'' \emph{arXiv preprint
  arXiv:2005.00959}, 2020.

\bibitem{cooley1965algorithm}
J.~W. Cooley and J.~W. Tukey, ``An algorithm for the machine calculation of
  complex {Fourier} series,'' \emph{Mathematics of computation}, vol.~19,
  no.~90, pp. 297--301, 1965.

\bibitem{shady2019correction}
S.~Abu~Hussein, T.~Tirer, and R.~Giryes, ``Correction filter for single image
  super-resolution: Robustifying off-the-shelf deep super-resolvers,'' in
  \emph{IEEE Conference on Computer Vision and Pattern Recognition (CVPR)},
  2020.

\end{thebibliography}

\end{document}